\documentclass[twoside,twocolumn]{article}
\usepackage{blindtext} 
\usepackage[sc]{mathpazo}
\usepackage[T1]{fontenc}
\usepackage{microtype}
\usepackage{graphicx}
\usepackage[english]{babel} 
\usepackage[hmarginratio=1:1,top=32mm,columnsep=20pt]{geometry} 
\usepackage[bf, small]{caption}
\usepackage{booktabs}

\usepackage{enumitem}
\setlist[itemize]{noitemsep}

\usepackage{abstract}
\usepackage{titlesec} % Allows customization of titles
\renewcommand\thesection{\Roman{section}} % Roman numerals for the sections
\renewcommand\thesubsection{\roman{subsection}} % roman numerals for subsections
\titleformat{\section}[block]{\large\scshape\centering}{\thesection.}{1em}{} % Change the look of the section titles
\titleformat{\subsection}[block]{\large}{\thesubsection.}{1em}{} % Change the look of the section titles

\usepackage{fancyhdr} % Headers and footers
\usepackage{titling} % Customizing the title section

\usepackage{hyperref} % For hyperlinks in the PDF

\usepackage{dcolumn}% Align table columns on decimal point
\usepackage{bm}% bold math
\usepackage[utf8]{inputenc}
\usepackage[T1]{fontenc}
\usepackage{mathptmx}

\pretitle{\begin{center}\Large\bfseries} % Article title formatting
\posttitle{\end{center}} 
\title{Comparative analysis for meaningful interpretation of rare-earth oxide M$_{4,5}$ energy loss edges} 
\author{%
\textsc{Ellis Kennedy$^1$, Christina Choi$^2$, M.C. Scott$^{1,2}$} \thanks{mary.scott@berkeley.edu} \\[.8ex] 
\small $^1$Department of Materials Science and Engineering, University of California Berkeley, Berkeley,\\ \small California, 94720, USA \\ \small $^2$NCEM, Molecular Foundry, Lawrence Berkeley National Laboratory, Berkeley, California, 94720, USA }

\date{\today} % Leave empty to omit a date
\renewcommand{\maketitlehookd}{%
\begin{abstract}
\noindent The magnetic, electronic, and optical properties of rare earth-oxides are directly influenced by the valency of the metallic cation. With the development of next generation electron energy-loss spectrometers, high-energy lanthanide fine structure can be studied with improved signal-to-noise for quantitative analysis. Unfortunately, the behavior of rare-earth \textit{4f} orbital electrons is not well understood. To establish best practices for analysis of energy-loss spectra from lanthanide oxides, we have performed a comparative study of the  four traditional white line analysis methods extended to lanthanide M$_{4,5}$ edges resulting from \textit{3d}$\rightarrow $\textit{4f} orbital transitions using data from Gatan’s \textit{EELS Atlas}. The M$_4$/M$_5$ spectral feature ratios were examined as a function of \textit{4f} occupancy. The M$_4$/M$_5$ spectral feature ratio decreases exponentially as \textit{4f} occupancy increases, except for a plateau between Sm$^{3+}$ and Dy$^{3+}$. The full-width at half the maximum intensity of the M$_4$ edges shows increased broadening for Sm$^{3+}$ through Dy$^{3+}$. We suggest that the plateau results from \textit{4f}-orbital half-filling and is explained through the relationship between electron transition probability and transition lifetime as expressed through Fermi’s Golden Rule.  Of the four spectral analysis methods described, only the integrated area method can be ascribed a quantitative physical interpretation.
\end{abstract}
}

\begin{document}

\bibliographystyle{ieeetr}

\maketitle

\section*{Introduction}

Sharp ionization peaks known as “white lines” are some of the most distinctive and useful features in the core-loss region of electron energy-loss spectroscopy (EELS). These edges result from the excitation of \textit{2p} core electrons to unoccupied d-states near the Fermi level in the transition metals and the excitation of \textit{3d} electrons to f-states near the Fermi level in the RE elements \cite{manoubi_quantitative_1990, pearson_measurements_1988,zaanen_l_23_1985}. The sharp edges aid in the determination of chemical species and analysis of their electron valencies. The L$_{2,3}$ white lines visible in transition metal spectra are well documented \cite{graetz_white_2004, kurata_electron-energy-loss_1993, manoubi_quantitative_1990,pearson_white_1993,yedra_oxide_2014} and many methods for edge analysis are presented across the literature \cite{botton_quantification_1995, kurata_electron-energy-loss_1993, riedl_extraction_2006}. However, information on the M$_{4,5}$ white lines present in RE spectra is sparse and generalized due challenges collecting high resolution spectral data in the high-energy EELS regime \cite{maclaren_eels_2018}.

The emergence of next generation EELS spectrometers that can better detect high-loss energy edges with improved signal-to-noise will enable improved spectral collection \cite{hart_direct_2017, maclaren_chapter_2019}. The commercialization of these spectrometers grants opportunity for in-depth analysis of near-edge fine structure in the high-loss regime. In anticipation of improved acquisition of high-loss EELS spectra that include RE M$_{4,5}$ edges, the extension of analytical techniques used for L$_{2,3}$ quantification to RE M$_{4,5}$ edges must be evaluated. A robust method for measuring \textit{4f} occupancy is beneficial for understanding the unique properties of binary RE oxides. 

While L$_{2,3}$ white line ratios are used for determining the valency of transition metal ions \cite{graetz_white_2004,yuan_electron_1999}, the relationship between valency and edge geometry requires further theoretical work \cite{wang_eels_2000}. The valency of RE ions in binary oxides is often even more difficult to determine and influenced by the initial electron state degeneracy \cite{gasgnier_formation_1986, manoubi_quantitative_1990}. Multiple valencies can exist simultaneously in a single sample \cite{kimura_mixed_1996, mott_rare-earth_1974, sankar_xanes_1983} and X-ray diffraction studies have shown that RE-oxides can be polyvalent or nonstoichiometric due to oxygen site vacancies \cite{kang_prediction_1998, manoubi_quantitative_1990}. X-ray absorption (XAS) and X-ray photoelectron spectroscopy (XPS) studies conclude that \textit{3d}→\textit{4f} initial electron energy multiplicity makes the complex structure of the RE series difficult to fully characterize. In XAS, only intensity from energy multiplets following dipole selection rules are visible, while in XPS all energy transition effects are visible \cite{crecelius_core-hole_1978, thole_3d_1985}. XAS studies on transition metals illustrate the need to account for crystal field splitting and cubic symmetries when focusing on electron-energy transition contributions to L$_{2,3}$ white lines \cite{de_groot_2p_1990}. While disparate, analogous spectral features are present in both x-ray methods and they can be analyzed following similar physical principles. Due to limited lanthanide EELS studies, attempts to reconcile EELS measurements with those of XAS and XPS have been limited. 

The relationship between the transition of electrons to the unoccupied \textit{4f}-orbital and the magnetic, optical, and electronic properties of RE alloys often depend on RE valency \cite{jeffries_degree_2010, marcinow_rare_1981, shenoy_electronic_2008}. Spectroscopic studies of RE alloys concentrate on the excitation of \textit{3d} core electrons to the \textit{4f} orbital  (\textit{3d}$^{10}$\textit{4f}$^n$→\textit{3d}$^9$\textit{4f}$^{n+1}$), which is the origin of many of the desirable material properties of RE alloys \cite{karnatak_profiles_1981, manoubi_quantitative_1990, thole_3d_1985}. 

To aid in future efforts to quantify RE EELS spectra and relate them to X-ray methods, we used data from the Gatan EELS Atlas to perform a comparative study of common white line analysis methods (Table  \ref{fig:tab1}) \cite{gatan_inc_eels_nodate}. We focused on the limitations, sensitivities, and physical interpretations of each method. We compared the results of common analysis methods when applied to the lanthanides (Fig. \ref{fig:fig1}). We denote the four methods, shown in Figure \ref{fig:fig2},  as (1) integrated M$_4$/M$_5$ edge area ratio with step function subtraction, (2) M$_4$/M$_5$ edge height ratio, (3) M$_4$/M$_5$ edge height ratio with step function subtraction, and (4) M$_4$/M$_5$ edge second derivative ratio. Multiplicity of initial \textit{3d} electron energy states is accounted for in two of the methods with the subtraction of step function of height ratio 3:2 from the M$_4$ and M$_5$ edges, respectively. Of the four white line analysis methods, only method (1) provides information on the complexities of the energy-loss interactions that occur during a \textit{3d}→\textit{4f} electron event. However, echoing conclusions from research on the L$_{2,3}$ edges, we caution against deriving solid state interpretations from M$_{4,5}$ white lines without a thorough analytical approach \cite{tan_oxidation_2012}.

\begin{table}[h]
\centering
\includegraphics[width=0.46\textwidth]{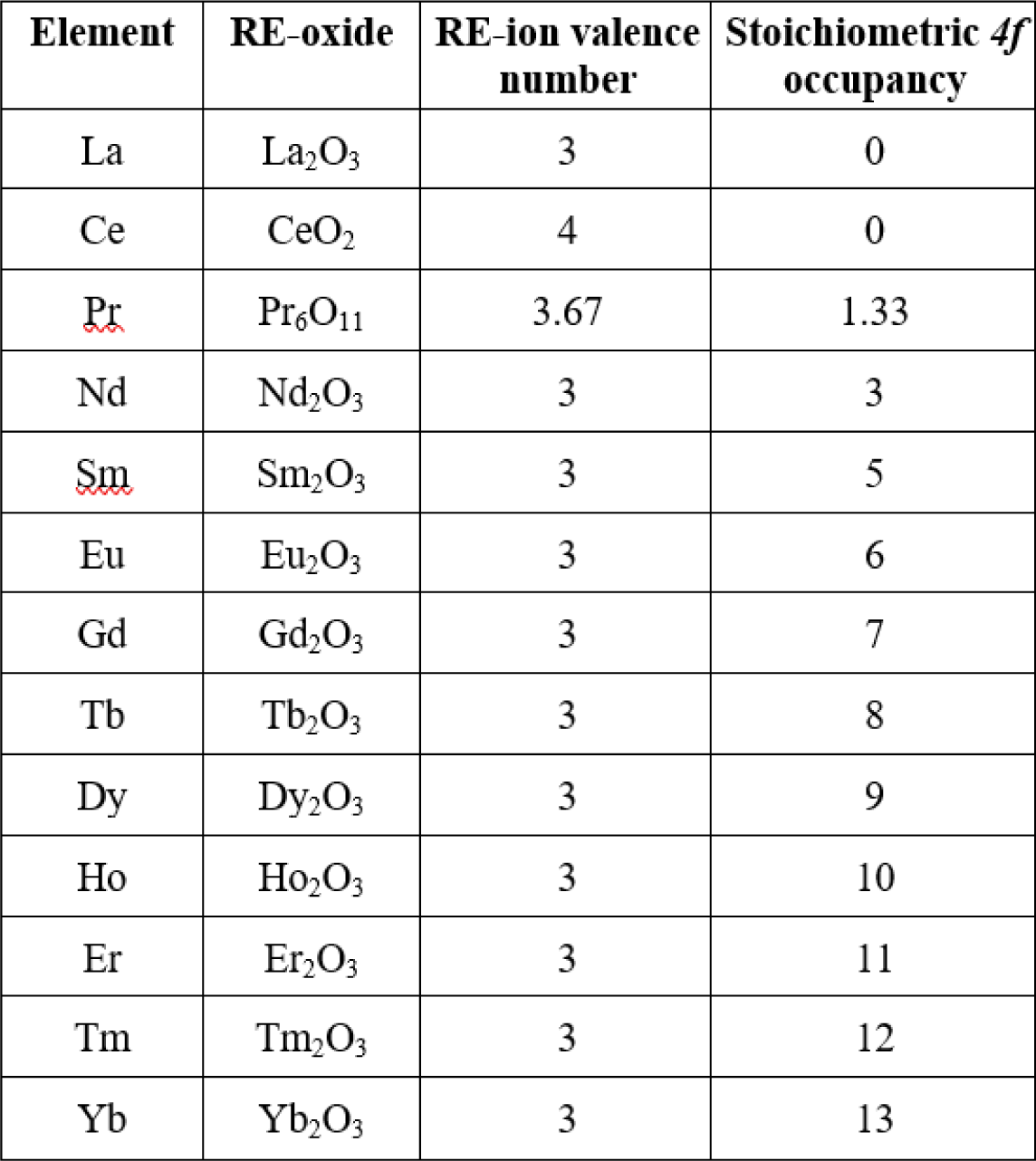}
\setlength{\belowcaptionskip}{0pt}
\caption{\label{fig:tab1} RE elements with corresponding oxides, ion valence numbers, and \textit{4f}-electron occupancy}
\end{table}

Through a comparison of the four white line analysis methods, we discuss the Sm-Dy plateau feature that is centered around Gd$^{3+}$, which breaks the exponential decreasing trend for M$_4$/M$_5$ intensity as a function of \textit{4f}-occupancy. Previous lanthanide oxide EELS studies do not offer an explanation for the sudden increase in M$_4$/M$_5$ spin-orbit feature ratio \cite{fortner_chemistry_1996}. The full width at half maximum intensity (FWHM) of each M$_{4,5}$ edge was measured and considered with regard to the physics governing electron transition lifetime. The loosely bound and delocalized nature of \textit{4f} electrons in the lanthanides allows them to more easily hybridize with states centered around oxygen atoms \cite{duan_hybridization_2004, zhang_strong-correlated_2018}. The behavior of the \textit{4f}-valence electrons is not fully understood, complicating analysis of lanthanide EELS spectra. We attribute the Sm-Dy plateau to energy-minimizing half-filling of the \textit{4f} orbital and \textit{5d} hybridization leading up to and beyond the Gd$^{3+}$ ion, which is expected to have a half-filled \textit{4f}-orbital. The Sm-Dy plateau is most prominent for method (1), the only method that reflects the probability distribution of an electron transition to an excited state.

%------------------------------------------------

\section*{Materials and Methods}

RE-oxide M$_{4,5}$ white lines were obtained from the Gatan EELS Atlas. High-energy regime M$_{4,5}$ data for elements 57-60 (La, Ce, Pr, Nd) and 62-70 (Sm, Eu, Gd, Tb, Dy, Ho, Er, Tm, Yb) were available (Table \ref{fig:tab1}). The edges lie between 832 and 1576 eV for this subset of the lanthanides. Data for Pm, which lacks a stable oxide, was absent. The same spectral data for each RE element were analyzed using four methods described in literature on white line analysis. We assumed negligible plural scattering since all spectra report a ratio of inelastic energy-loss intensity to zero-loss peak intensity below 1.0. The I$_{inelastic}$/I$_{ZLP}$ ratio for each sample is reported in the Gatan EELS Atlas and included in the Supplemental information with thickness range calculations for each sample \cite{ahn_inner_1985,gatan_inc_eels_nodate}. Ideally, the EELS spectra would be deconvolved with the zero-loss peak, but the required data were unavailable and the samples were determined to be sufficiently thin. All serial EELS spectra were collected under the following experimental parameters: a 200 keV operating voltage, a 100 mrad collection semi-angle, 20-40 probe sweeps, and 20-30 ms dwell times \cite{ahn_inner_1985}. 

\begin{figure}[h!]
\centering
\includegraphics[width=0.46\textwidth]{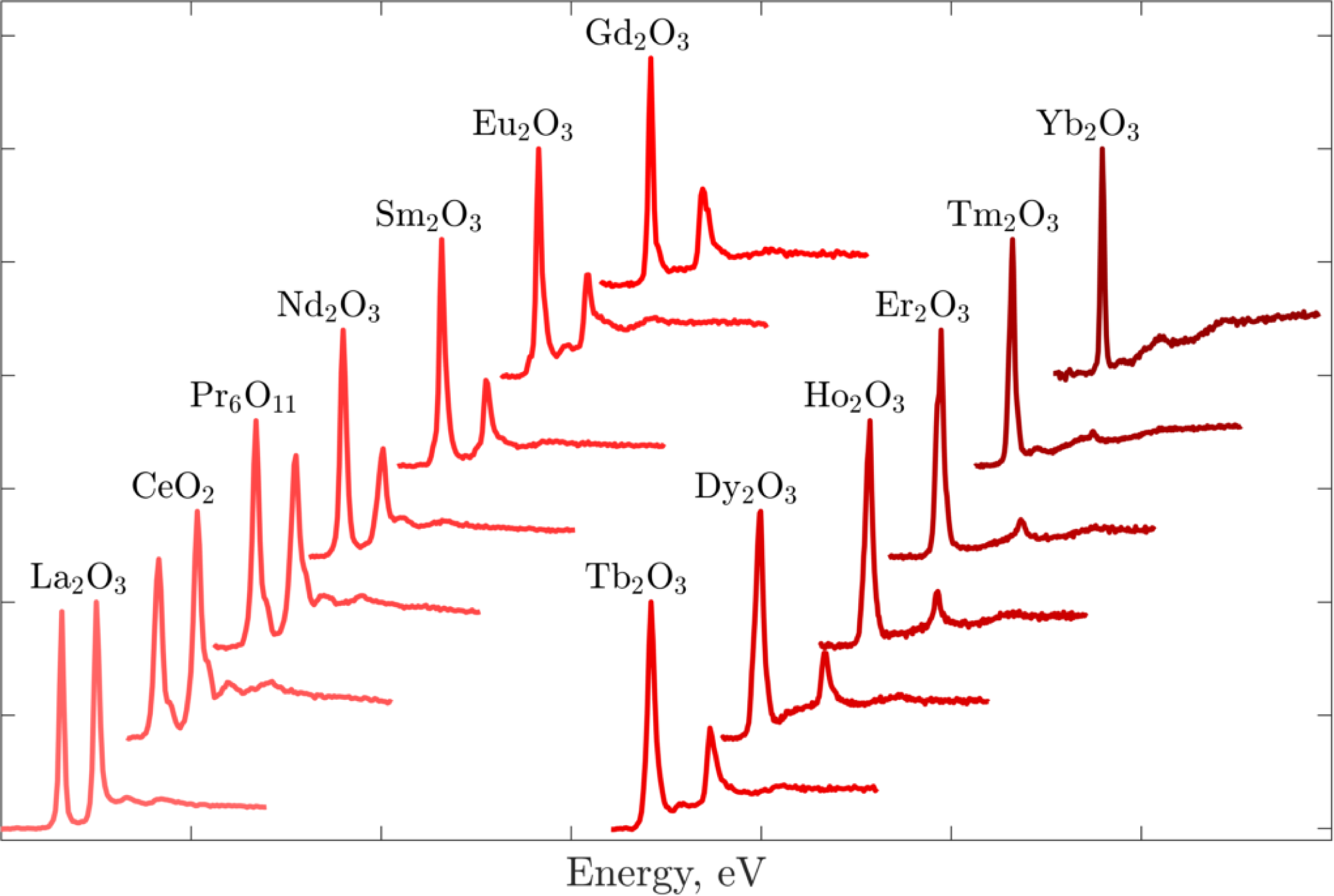}
\setlength{\belowcaptionskip}{0pt}
\caption{\label{fig:fig1} RE-oxide EELS spectra obtained from Gatan EELS Atlas and used to compare white line analysis methods \cite{gatan_inc_eels_nodate}. Both axes are in arbitrary units as the figure is intended to show relative trends in M$_4$ /M$_5$ edge feature ratio and in changing edge geometries with increasing RE Z-number and \textit{4f}-valency.
}
\end{figure}

A power law background subtraction was performed using the CSI Spectrum Analyzer plugin for ImageJ \cite{cueva_data_2012}. Background windows were 50 eV wide and ended at least 10 eV before the onset of the M$_5$ edge. For the M$_4$/M$_5$ edge second derivative ratio method, no background subtraction was applied because it yields similar results as when the background is correctly subtracted, making this method more robust against inaccurate background assignment.

We applied four methods used to quantify L$_{2,3}$ white lines to the M$_{4,5}$ edges of the lanthanide series. For each method, we observed the ratio between a specific M$_4$  and M$_5$ spectral edge feature as a function of \textit{4f} valence orbital filling in binary RE oxides. The methods are shown in Figure \ref{fig:fig2} and detailed in the Supplemental information. All EELS spectra were sourced from the online Gatan EELS Atlas and are shown in Figure \ref{fig:fig1} \cite{ahn_inner_1985}. Table 1 includes the RE-oxides studied and their stoichiometric valencies and \textit{4f}-occupancies.

\begin{figure}[h]
\centering
\includegraphics[width=0.46\textwidth]{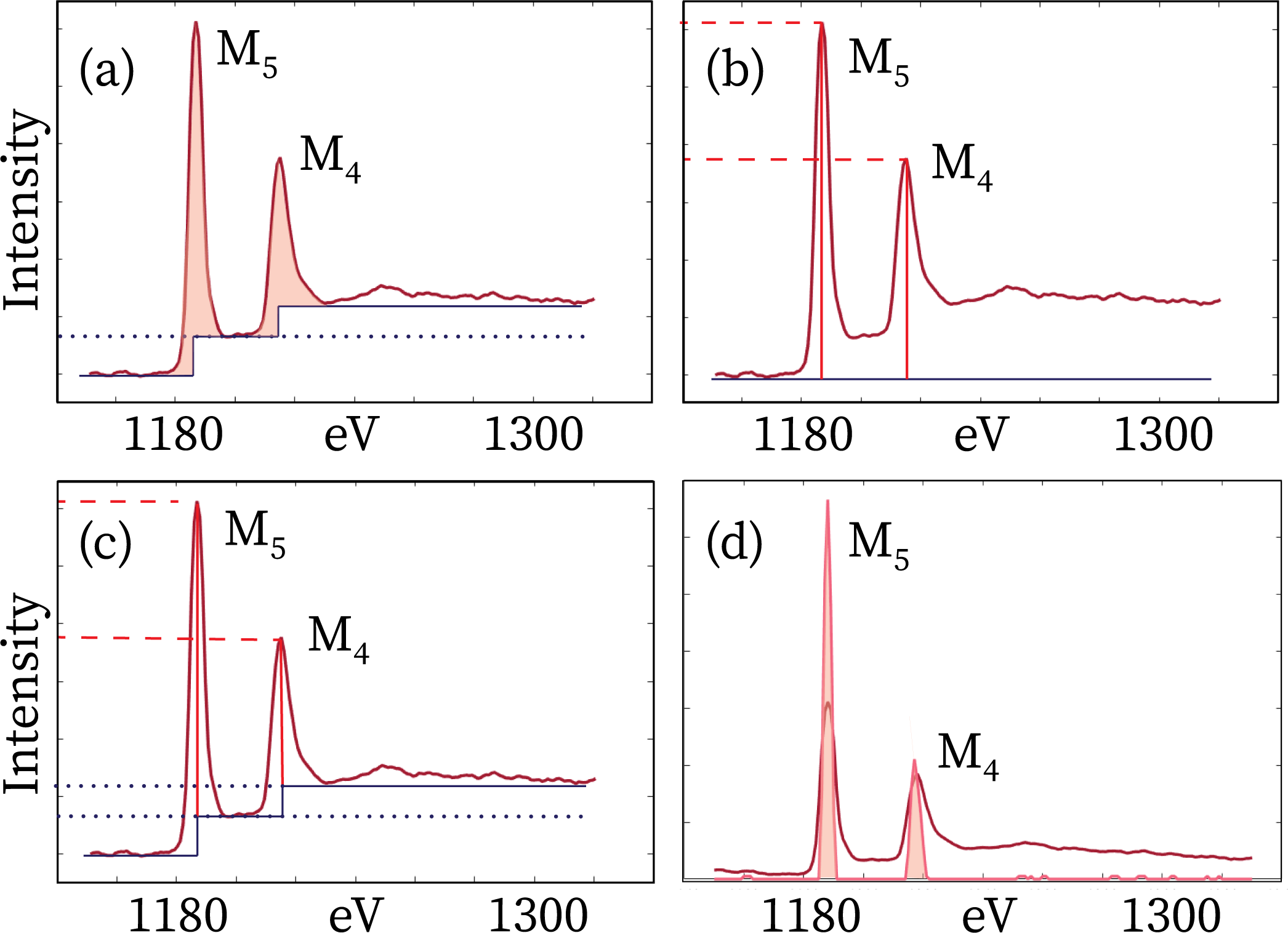}
\setlength{\belowcaptionskip}{0pt}
\caption{\label{fig:fig2} Illustration comparing the four methods of white line analysis that were used to study RE-oxide spectra. Gd$^{3+}$ M$_{4,5}$ edges from a Gd$_2$O$_3$ sample are shown as representative white lines. (a) Method (1), integrated area ratio with step function subtraction with the area of each edge shaded in. (b) Method (2), height ratio with intensities indicated on the vertical axis with dashed red lines. (c) Method (3), height ratio with step area subtraction with the edge intensities indicated on the vertical axis with dashed red lines and the subtracted step values indicated with dotted lines. (d) Method (4), the positive component of the second derivative ratio is shown in pink with the integrated area for finding M$_4$/M$_5$ ratio shaded in. The second derivative plot is overlain on the Gd$_2$O$_3$ spectra without background fitting.
}
\end{figure}

\section*{Results}
\begin{figure*}[h!]
\centering
\includegraphics[width=.58\textwidth]{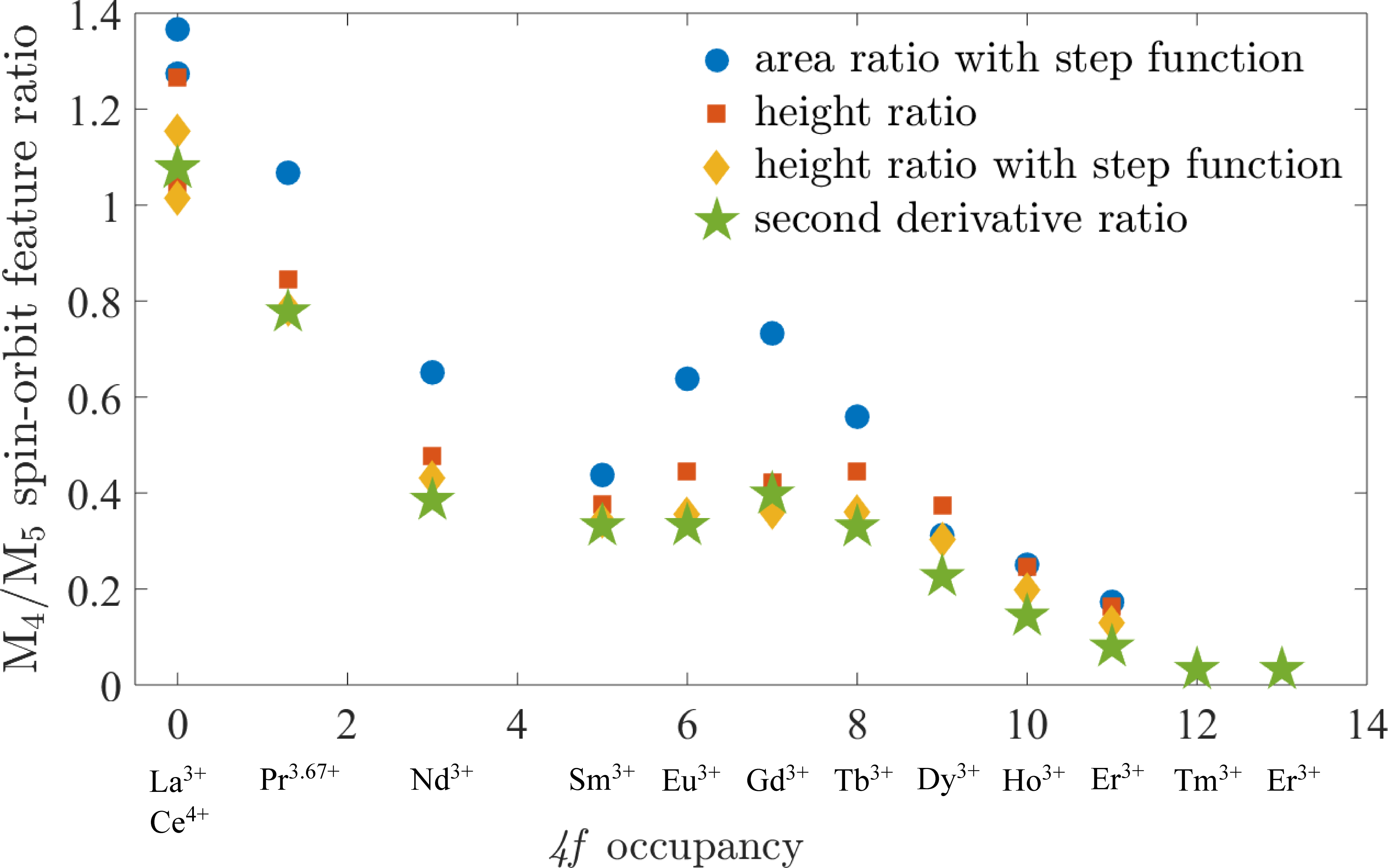}
\caption{\label{fig:fig3}Dependence of the M$_4$ /M$_5$ spin-orbit edge feature ratio on the stoichiometric electron occupancy of the \textit{4f}-orbital for the methods: (1) integrated edge area ratio with step function subtraction (blue circle), (2) edge height ratio (orange square), (3) edge height ratio with step function subtraction (yellow diamond), and (4) edge second derivative ratio (green star).}
\end{figure*}

The four spectrum analysis methods produce the M$_4$/M$_5$ spin-orbit feature trends shown in Figure \ref{fig:fig3}, in which the white line feature ratio is plotted as a function of \textit{4f}-orbital filling. The four methods are roughly in agreement. The lighter RE-oxides have larger M$_4$/M$_5$ white line feature ratios than the heavier lanthanides. M$_4$/M$_5$ feature ratios initially decrease, but with diminishing slope until reaching an inflection point at a \textit{4f} occupancy of approximately five electrons (Sm$^{3+}$). At this occupancy, the M$_4$/M$_5$ feature ratio increases across all methods. The trend then plateaus for methods (2) and (3). For methods (1) and (4), the trend reaches a second maxima at seven electron \textit{4f} occupancy (Gd$^{3+}$). The M$_4$/M$_5$ feature ratio begins to decrease again past a \textit{4f} occupancy of eight electrons. Overall, the feature intensity of the M$_4$ energy-loss edges is similar to that of the M$_5$ in the early lanthanides, but its relative intensity decreases rapidly toward the later lanthanides. This is also apparent in Figure \ref{fig:fig1}.

Comparing the four methods, the M$_4$/M$_5$ feature ratio as a function of \textit{4f}-occupancy of method (1) deviates most significantly from the other three methods. The M$_4$/M$_5$ feature ratio for method (1) is, in general, higher than that for the other methods. The Sm-Dy plateau is most prominent when the data is analyzed with this method. There is a significant increase in the M$_4$/M$_5$ feature ratio centered about Gd$^{3+}$ with seven electrons in the \textit{4f}-orbital. The other three methods display a less exaggerated Sm-Dy plateau centered at Gd$^{3+}$. As it is the integrated areas beneath white lines that relate to electron transition probability, method (1) is the only measurement with physical significance.

FWHM measurements of the M$_{4,5}$ white lines relate the energy spread of the white lines. FWHM values for the post-background subtraction M$_{4,5}$ edges are plotted in Figure \ref{fig:fig4}. The M$_4$ edge FWHM decreases for RE-oxide ions with \textit{4f}-occupancies above that of Pr$^{3.67+}$  until reaching Eu$^{3+}$ with six \textit{4f} electrons. There is a peak in the M$_4$ FWHM trend at Gd$^{3+}$, where the \textit{4f}-orbital is half-filled. The FWHM then decreases again until spiking at Er$^{3+}$  with eleven \textit{4f} electrons. 

Similar to the M$_4$ FWHM trend, the M$_5$ FWHM trend initially decreases for \textit{4f}-occupancies above that Pr$^{3.67+}$. Contrasting the trend in M$_4$ FWHM, the M$_5$ FWHM continues to decrease with a minimum at Gd$^{3+}$. Between Sm$^{3+}$ and Dy$^{3+}$, the M$_4$ and M$_5$ FWHM trends are inverses, both with extrema centered at Gd$^{3+}$. The increase in area under the M$_4$ edge around Gd$^{3+}$ and the accompanying decrease in area under the M$_5$ edge both contribute to the prominence of the Sm-Dy plateau shown in Figure \ref{fig:fig3}, with the broadening likely attributable to \textit{4f} orbital half-filling. If significant multiple scattering were to occur it would disproportionately broaden the M$_4$ edge. However, because the  ratio of inelastic energy-loss intensity to zero-loss peak intensity is below 1.0 for all spectra, the samples are sufficiently thin for plural scattering to be considered negligible \cite{ahn_inner_1985}.

%------------------------------------------------

\section*{Discussion}

\textit{Electron initial state multiplicity}\\
Measured M$_{4,5}$ edges arise from a transition from an initial state in the d orbitals to a final state in the f orbitals. To account for initial state electron multiplicity in this scenario, we extend the same physical principles of multiplicity used for analyzing L$_{2,3}$ edges to the M$_{4,5}$ edges because of similarities in electron exchange interactions and well-defined edge separation \cite{krivanek_elnes_1990,pearson_white_1993, wang_eels_2000}. When this method has been applied to the L$_{2,3}$ transition metal white lines, the appropriate ratio of the step is 2:1, which correctly accounts for the multiplicity of the initial \textit{2p} electron states (four \textit{2p}$^{3/2}$ and two \textit{2p}$^{1/2}$). In theory, the step height ratio for the RE M$_{4,5}$ white lines is 3:2 because of the six initial \textit{3d}$^{5/2}$ electron states and the four initial \textit{3d}$^{3/2}$ electron states, as shown in Figure \ref{fig:fig5}. The splitting of initial electron states results in the separate M$_4$ (\textit{3d}$^{3/2}$$\rightarrow $\textit{4f}$^{5/2}$) and M$_5$ (\textit{3d}$^{5/2}$$\rightarrow $\textit{4f}$^{7/2}$) edges observed in EELS \cite{thole_3d_1985}. In practice, the trough between the M$_5$ and M$_4$ edges often dips beneath the height of the lower step when the 3:2 ratio is applied from the continuum. When this happens, the energy at the nadir is taken as the height of the lower step. The step function is subtracted from the M$_{4,5}$ spectra for methods (1) and (3) with the implication that the energy multiplicity embedded in the continuum is accounted for. 

\textit{Comparison of analysis methods}\\
The four methods for EELS analysis under comparison have different sensitivity to spectral features in EELS data. Considerations for interpreting lanthanide EELS data analyzed using these methods are outlined in detail in the Supplemental information. 

\begin{figure}[h!]
\includegraphics[width=0.46\textwidth]{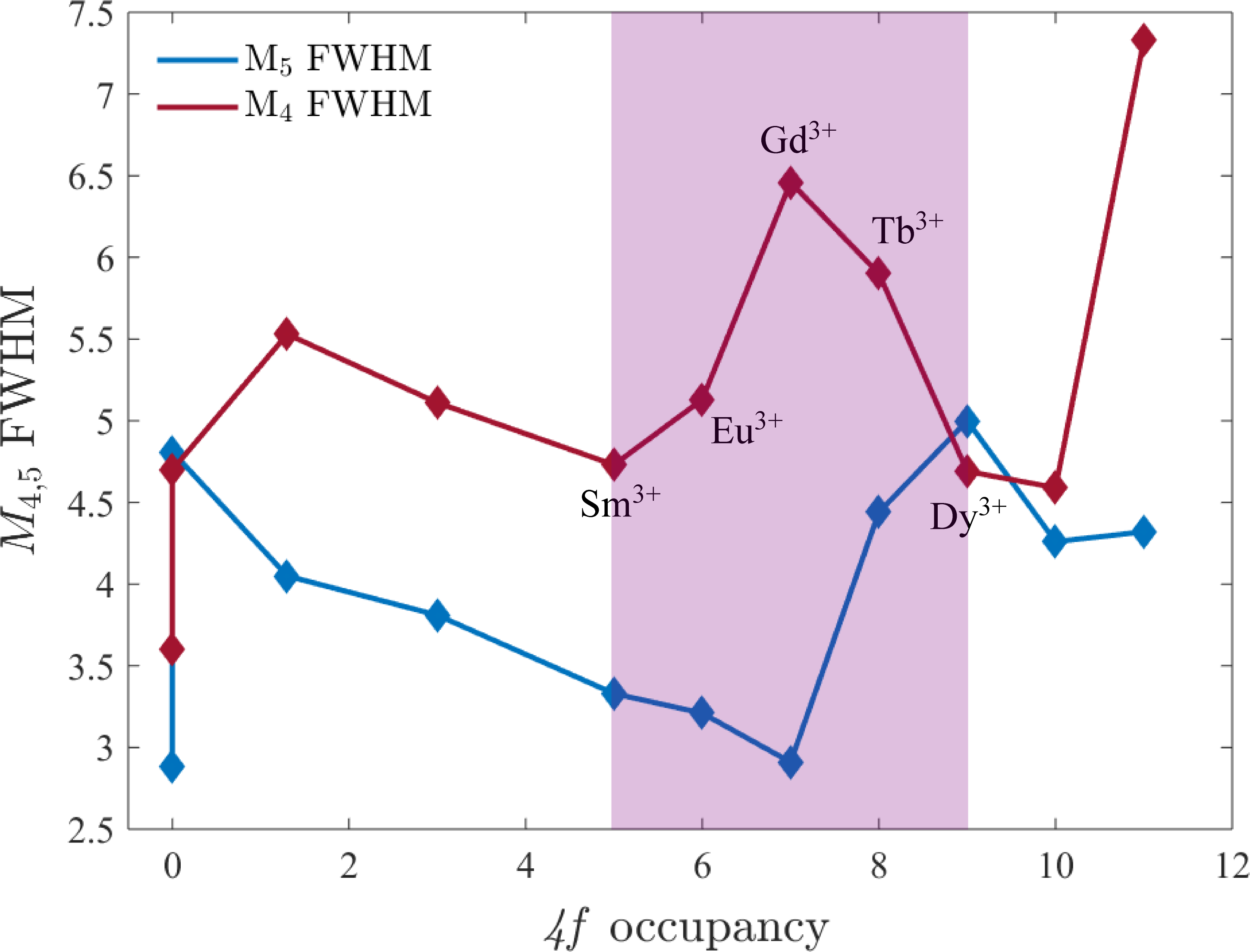}
\setlength{\belowcaptionskip}{0pt}
\caption{\label{fig:fig4} Measured FWHM of the M$_4$ (red) and M$_5$ (blue) edges for \textit{4f}-occupancies of elements 57-60 and 62-68.  The FWHM of M$_4$ has a maximum at Gd$^{3+}$, the half-filled state. The FWHM of M$_5$ is minimized at Gd$^{3+}$. The shaded region is centered about a \textit{4f}-occupancy of seven electrons and indicates the RE oxides that contribute to the enhanced Sm-Dy plateau observed with method (1) in Figure \ref{fig:fig1}.
}
\end{figure}

\begin{figure}[h!]
\centering
\includegraphics[width=0.42\textwidth]{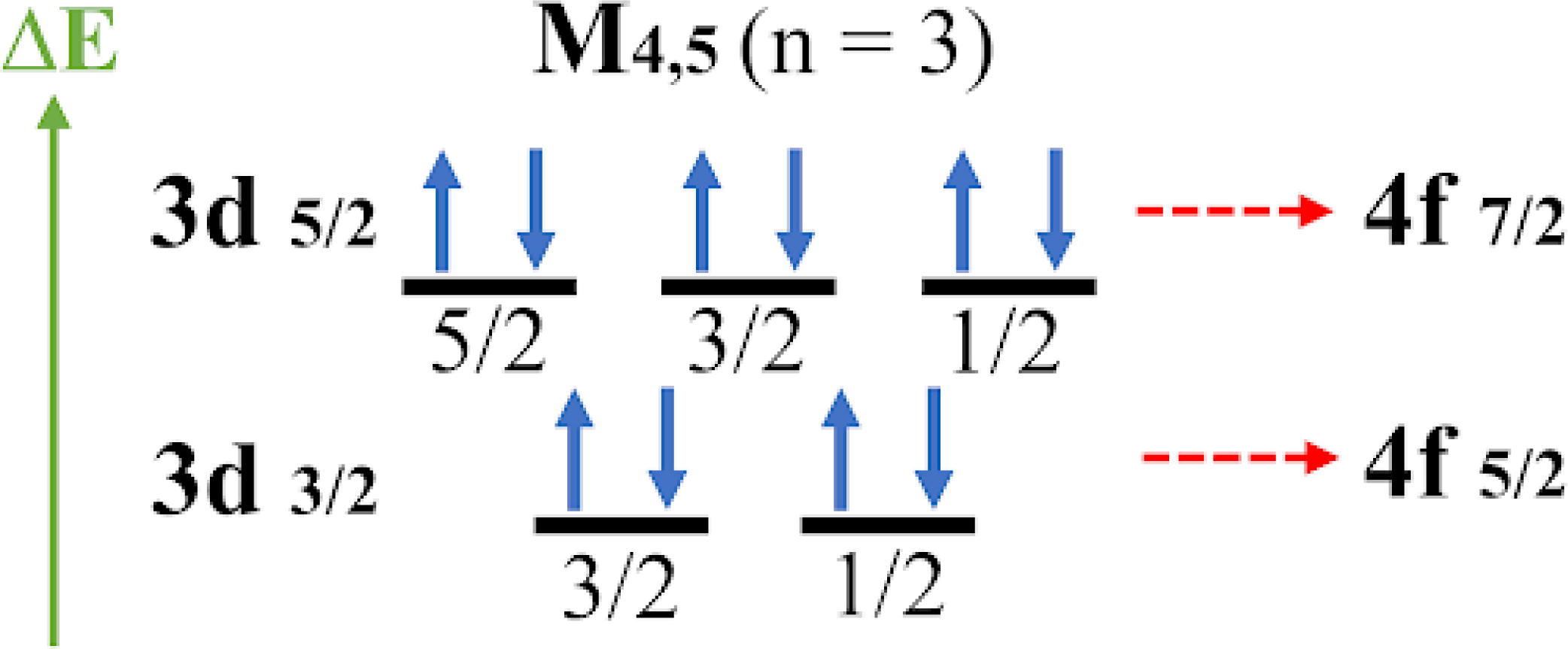}
\setlength{\belowcaptionskip}{0pt}
\caption{\label{fig:fig5}Spin orbit splitting diagram displaying the origin of M$_{4,5}$ edges and the 3:2 ratio of the \textit{3d}$^{5/2}$ and \textit{3d}$^{3/2}$ electron states.
}
\end{figure}

\textit{Physical interpretation of lanthanide EELS data}\\
For M$_{4,5}$ transitions, the initial energy states are the \textit{3d} orbitals and the final states are the \textit{4f} orbitals. As indicated in Figure \ref{fig:fig5}, the \textit{3d}$^{3/2}$ initial energy state is of lower energy than the \textit{3d}$^{5/2}$ initial state. An increasing M$_4$/M$_5$ ratio indicates that the energy-loss signal is increasingly composed of \textit{3d}$^{3/2}$$\rightarrow $ \textit{4f}$^{5/2}$ transitions. According to Hund’s rule, unpaired electrons will fill available \textit{4f} orbitals first, with unpaired electrons filling \textit{4f}$^{7/2}$ orbitals beginning with Sm$^{3+}$. This corresponds to the first inflection point in Figure \ref{fig:fig4}, as this represents when fewer  final states begin to be available for the M$_4$ transition. This trend continues until Tb$^{3+}$, where electron pairs begin to form and fewer final states are available for the M$_5$ transition \cite{benelli_magnetism_2002}. As the \textit{4f} orbitals are occupied, transitions to the lower energy \textit{4f}$^{5/2}$ states are blocked by the Pauli exclusion principle and the loss spectra are dominated by excitations to the \textit{4f}$^{7/2}$ states \cite{williams_transmission_2009}. The empty, half full, and completely full states of the \textit{4f}-orbital provide the greatest energetic stability \cite{boyd_quantum_1984}. It is possible that to increase electron configuration stability, the trivalent RE ions of Sm$^{3+}$ through Dy$^{3+}$ have half filled or nearly half filled \textit{4f}-orbitals. The elements surrounding Gd$^{3+}$ would also have approximately seven \textit{4f} electrons if this is the case and be partially \textit{5d} hybridized\cite{duan_hybridization_2004, komesu_4f_2009}. 

If the trivalent RE ions adjacent to Gd$^{3+}$ in Figure \ref{fig:fig3} all have nearly seven \textit{4f}-electrons, then the similarity in M$_4$/M$_5$ ratios between Sm$^{3+}$, Eu$^{3+}$, Gd$^{3+}$, Tb$^{3+}$, and Dy$^{3+}$ is explained. However, this does not explain why the Sm-Dy plateau rises in the edge area ratio method.  The \textit{3d}$^{10}$\textit{4f}$^n$$\rightarrow $\textit{3d}$^9$\textit{4f}$^{n+1}$ electron energy transition probability is governed by the density of available final states and the pathways to the final states.  The M$_4$ Sm-Dy plateau white lines exhibit broadening and increased relative integrated areas compared to the trend observed in the rest of the lanthanides. Through Heisenberg’s uncertainty principle, the increase in energy indicates lower mean transition lifetimes and higher transition probabilities and rates \cite{williams_transmission_2009}. Fermi’s Golden Rule describes the dependence of transition probability on density of final states and coupling between the final and initial states, and can explain the differences in the intensities of the M$_{4,5}$ spectral lines \cite{merzbacher_quantum_1998}. According to Fermi’s rule, increased transition probabilities, and therefore increased integrated area, should occur when more final states are available and coupling between states is higher. 

An increase in the area under the M$_4$  edge is caused by a greater  probability of electron transition to an excited state, which is indicative of an increase in the density of final states or in the accessibility of these final states. Stabilizing orbital half-filling increases both the density of final states and the transition pathways by offering more routes from initial state to final state. This indicates that the mean \textit{4f} occupancy is greater than stoichiometrically expected for Sm$^{3+}$ and Eu$^{3+}$ and less than expected for Tb$^{3+}$ and Dy$^{3+}$. The M$_5$ Sm-Dy white lines exhibit reduced integrated areas, indicating reduced electron transition rates to \textit{4f}$^{7/2}$ final states. The Sm-Dy plateau is not only explained by half-filling and hybridization, but is expected based on Fermi’s rule. 

\textit{Limitations in interpretation of EELS data for lanthanide oxides}\\
The covalent character of RE-oxides limits the relationship between the cation oxidation state and the stoichiometric \textit{4f}- occupancy \cite{tan_oxidation_2012,maclaren_chapter_2019}. There are additional limitations to trusting stoichiometry and RE-oxygen atomic ratios may differ from those presented in Table \ref{fig:tab1}. For example, La$^{4+}$ and Ce$^{3+}$ are both expected to have a \textit{4f}-orbital occupancy of zero electrons. Only the second derivative method results in the M$_4$/M$_5$ ratio of La$^{4+}$ being slightly greater than that of Ce$^{3+}$. The other methods result in a greater M$_4$/M$_5$ ratio for Ce$^{3+}$, with the height ratio method resulting in the greatest difference. Studies have shown that RE-oxides often contain oxygen-site vacancies and that higher oxides, such as Pr$_6$O$_11$, frequently split into two or more polytypes in environmental conditions \cite{adachi_binary_1998, shafer_relationship_1972}. The existence of multiple oxides within a single sample would be difficult to parse out in most M$_{4,5}$ EELS data, particularly if one oxide dominates. Modern advances in EELS detectors may assist by providing the resolution required to make out fine spectral features in the high energy-loss regime.

Additionally, the possibility of \textit{4f}-\textit{4f} electron energy transitions and \textit{5d} hybridization contributing to the shape of the plots in Figure \ref{fig:fig3} should also be considered. The Judd-Ofelt theory provides a theoretical framework for analyzing RE \textit{4f}-\textit{4f} electron energy transitions and outlined successive splitting of energy levels into many schemes of energy degeneracy \cite{judd_optical_1962,ofelt_intensities_1962}. The energy splitting could influence the preferred \textit{3d}$\rightarrow $ \textit{4f} transitions of energy-loss electrons, but is likely less significant than the effects of \textit{4f} half filling and \textit{5d} hybridization for energy minimization.

Ultimately, the EELS spectral features and plateau feature observed for M$_4$/M$_5$ feature ratios as a function of \textit{4f}-occupancy result from many factors, including ions stabilized with half-filled \textit{4f} orbitals, the presence of multiple oxides in a single sample, oxygen-vacancies resulting in nonstoichiometric compositions, and \textit{5d} hybridization. It seems likely that in a system with such complex electron interactions, more than one (if not all) of these properties will contribute to M$_{4,5}$ edge geometry in the lanthanides. Determining the weighted contributions of each requires furthering theoretical knowledge on \textit{4f}-electrons and collecting well-resolved EELS data from standard lanthanide oxides.

\section*{Summary}

In conclusion, we hope to offer insight for those performing EELS analysis on the lanthanide M$_{4,5}$ edges. We used four methods of white line analysis to plot M$_4$/M$_5$ spectral feature ratios as a function of \textit{4f} occupancy for lanthanide oxides. For two of the methods, we extended principles of initial state electron multiplicity by subtracting a 3:2 energy-loss intensity step-function from the M$_5$ and M$_4$ edges, respectively, to account for the broken degeneracy of \textit{3d} electron energies in a crystal field. There is a general decrease in M$_4$/M$_5$ feature intensity as the \textit{4f} fill level increases, but the Sm-Dy plateau emerges toward the half-fill point that is centered about Gd$^{3+}$. The Sm-Dy plateau is most evident using method (1), indicating that the Sm-Dy plateau height is dependent on the area beneath the M$_{4,5}$ edges. M$_4$ edge broadening responsible for the Sm-Dy plateau is attributed to \textit{4f} orbital half-filling, but is not fully understood, as electron itinerancy and \textit{5d} hybridization may contribute to the electron interactions responsible for this feature.

Of the four methods described, the integrated area method is the only method that provides quantitative information on M$_{4,5}$ electron energy-loss behavior by incorporating the probability of electron transitions to excited states and the interactions of these excited states with other atomic states in the system. However, further theoretical development is required to explain edge geometry in terms of electron energy transitions and interactions and caution should be exercised when attributing electron-loss intensity to solid state effects. M$_{4,5}$ white line analysis does not provide an accurate measurement of cation oxidation state, especially not for RE-oxides with sizeable covalent character \cite{tan_oxidation_2012}. Density functional theory calculations of lanthanide oxides with varying \textit{4f}-occupancy would help explain ground state transitions and \textit{4f} electron interactions responsible for the M$_{4,5}$ edges observed with EELS. Those who desire simple fingerprinting can use the edge height method as it conveys a similar trend as the other methods and provides quick approximation, while the second derivative height ratio method provides analysis that is unbiased with respect to background. Researchers would benefit from updated, absolute-energy positioned RE-oxide spectra in the \textit{EELS Atlas} and accompanying zero-loss peak data.

With the emergence of next generation direct-detection detectors, low signal-to noise high-energy regime EELS data can be collected from the lanthanide series. A theoretically-supported understanding of the experimentally observed energy-absorption and energy-loss features encoded in M$_{4,5}$ EELS data would forward understanding of \textit{4f} electron behavior in the lanthanides.

\section*{Acknowledgments}
This work was  supported by the National Science Foundation, DMR-1548924. Work at the Molecular Foundry was supported by the Office of Science, Office of Basic Energy Sciences, of the U.S. Department of Energy under Contract No. DE-AC02-05CH11231. We thank Liam Spillane at Gatan Inc. for providing information on the samples from which the spectra were collected. We thank Rohan Dhall at the Molecular Foundry, Lawrence Berkeley National Laboratory for enlightening conversations that influenced the direction of the work.

%---------------------------------------------------

\bibliography{IEEEabrv,M45EELS}
\newpage

\section*{Supplemental information}

\begin{table}[h!]
\centering
\includegraphics[width=0.46\textwidth]{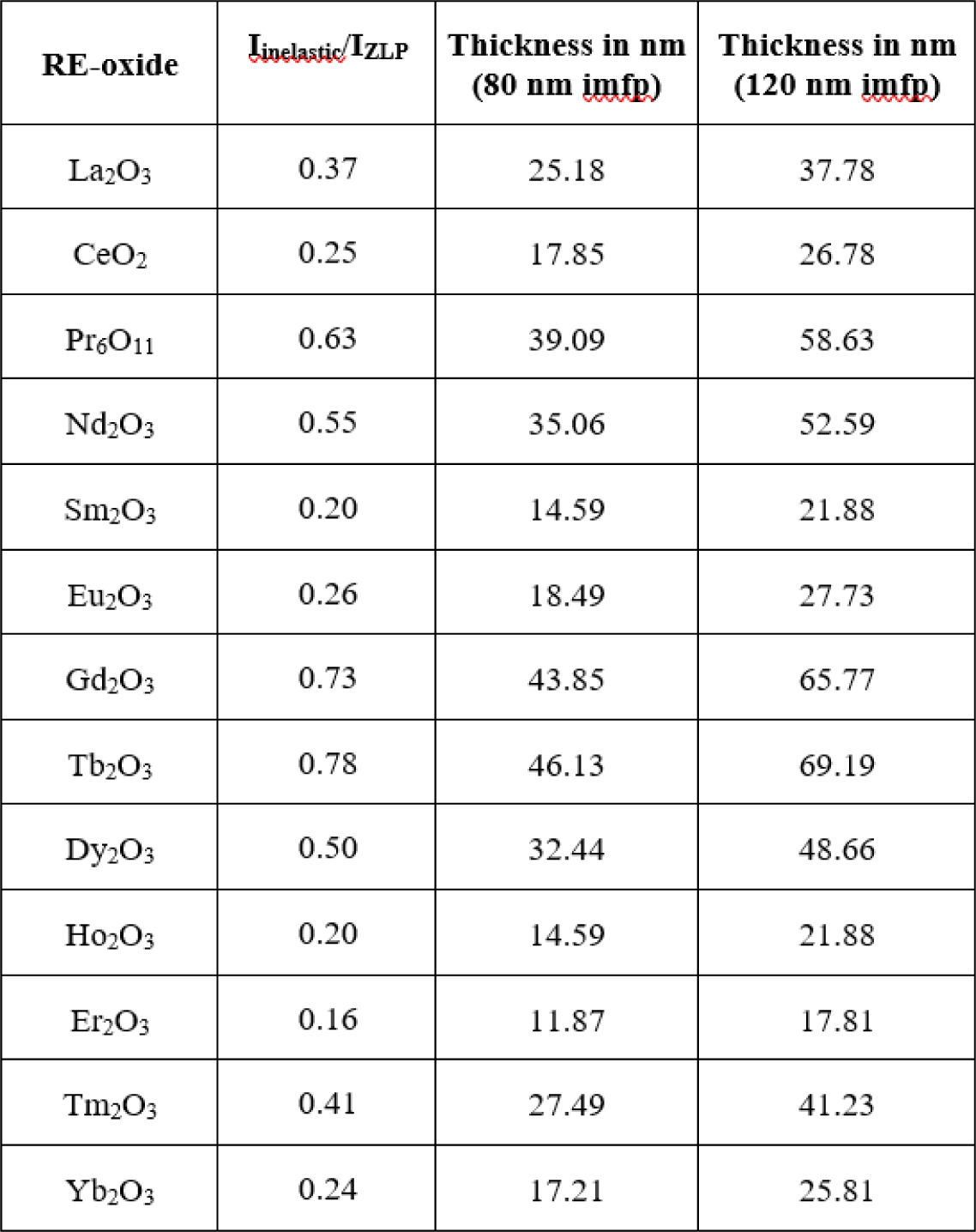}
\setlength{\belowcaptionskip}{0pt}
\caption{\label{fig:sup1}Estimates of RE-oxide  thickness based on lower (80 nm) and upper (120 nm) bounds of inelastic mean free path lengths (imfp) provided in the EELS Atlas (Ahn and Krivanek, 1983). All samples are thin films except Sm$_2$O$_3$, which was in nanoparticle form.}
\end{table}

\subsubsection*{Detailed description of methods for analyzing white line spectra}
\textit{Method 1: Integrated area ratio with step function subtraction}\\
Oxides of elements 57-60 and 62-68 were analyzed using the integrated edge area ratio with step function subtraction. Tm$^{3+}$ and Yb$^{3+}$ were excluded because significant M$_4$  edge broadening and reduced edge intensity made distinguishing the edge onset and shape from the background continuum subjective. After background subtraction, the data were normalized and a step function threshold was subtracted from the background. Using the method demonstrated by Pearson et al. 1988 \cite{pearson_measurements_1988}, the white line intensity ratio was determined with consideration for the multiplicity of the initial \textit{3d} electron states. A horizontal line was fit over a range of 50 eV immediately following the M$_4$ white line. 

A double step function was created from the horizontal line modeling the background continuum with the step onsets occurring at the M$_4$ and M$_5$ white line maxima eV values as shown in Figure \ref{fig:fig2}(a) on a representative spectrum of the Gd$^{3+}$ M$_{4,5}$ edges. Edge onsets were selected using those listed in the EELS Atlas and edge ends were determined manually as the energy at which the edge intensity coalesces into the background intensity. The integrated area beneath the M$_{4,5}$ edges and above the step function, indicated by light shading in Figure \ref{fig:fig2}(a), was used to determine the M$_4$/M$_5$ edge area ratio. 
 
\textit{Method 2: Height ratio}\\
A simple analysis of edge height ratio was performed on the same eleven RE-oxide samples. Like the L$_2$/L$_3$ edge height ratio, the M$_4$/M$_5$ edge height ratio is not generally considered a robust method for white line analysis because it fails to account for the range of transition energies present during ionization and assumes consistent white line width. It is used for oxide fingerprinting \cite{tan_oxidation_2012, wang_eels_2000}. This method is included for comparison against the other methods presented. For the eleven binary oxides analyzed, after background subtraction, the data were normalized and the relative maximum intensities, or edge heights, of the M$_4$ to the M$_5$ edges of each RE ion were calculated. This is shown in Figure \ref{fig:fig2}(b) with the edge maxima marked on the vertical intensity axis.

\textit{Method 3: Height ratio with step function subtraction}\\
To account for the effect of the multiplicity of the initial \textit{3d} electron states, we analyzed the edge height ratio after subtraction of a step function from each spectra. The step function subtraction procedure was the same as that carried out for determining the integrated area ratio. For the eleven binary oxides, after background subtraction, the data were normalized and a 3:2 step function was subtracted from the background. Then, the relative maximum intensities of the M$_4$ to the M$_5$ edges of each RE ion were calculated as shown in Figure \ref{fig:fig2}(c). Similar to method (2), this method assumes consistent white line width and is included for comparison with the other methods.

\textit{Method 4: Second derivative ratio}\\
Binary oxides of elements 57-60 (La, Ce, Pr, Nd) and 62-70 (Sm, Eu, Gd, Tb, Dy, Ho, Er, Tm, Yb) were analyzed using the edge second derivative ratio method. Unlike the other methods, this method does not require background subtraction because the positive components of the second derivative corresponding to the M$_{4,5}$ edges are pronounced and unlikely to be mistaken as part of the second derivative of the continuum \cite{martin_chemical_1996}. We used D.R.G. Mitchell’s script \textit{Measure EELS Peak Intensities} (ver. 4, published Apr. 2020) \cite{mitchell_scripting-customised_2005} to perform the second derivative calculation and found a similar trend to Fortner and Buck in the M$_4$/M$_5$ spin-orbit feature ratio (their Figure 1) \cite{fortner_chemistry_1996} with the inclusion of EELS Atlas data for La$^{3+}$, Ce$^{4+}$, and Pr$^{3.67+}$. 

The second derivative method yielded distinguishable peaks for the M$_4$ and M$_5$ edges of all RE-oxides analyzed, including Tm and Yb. Figure \ref{fig:fig2}(d) overlays Gd$^{3+}$ M$_{4,5}$ edges and the positive components of the second derivative of the M$_{4,5}$ edges over the same energy-loss region. By preventing the potential for slight differences in manual selection of the power law background window for subtraction and removing the need to approximate the edge onsets and ends, this method is robust against methodological inconsistencies. The integrated area of the M$_4$ and M$_5$ second derivative features, indicated by light shading in Figure \ref{fig:fig2}(d), were used to compute the M$_4$/M$_5$ second derivative ratio.

\subsubsection*{Detailed comparison of differences observed trends in the M$_4$/M$_5$ spin-orbit feature vs. \textit{4f}-occupancy with respect to method of data analysis}

\textit{Method 1: Integrated edge area ratio}\\
The integrated edge area ratio, method (1), is sensitive to total integrated intensities of the M$_{4,5}$ electron energy loss within an energy range defined by the edge width. The step function subtraction accounts for the multiplicity of \textit{3d} energy transitions by removing portions of the energy continuum beneath each edge in proportion to the number of degenerate initial states of the \textit{3d} electrons.  The M$_4$ edge maximum intensity decreases from left to right across the row. The M$_4$ becomes increasingly more difficult to separate from the background as the energy width increases and the edge height decreases. The FWHM measurements of the M$_{4,5}$ edges vary over the lanthanide oxides observed in this work. The trend of increased FWHM, indicating edge broadening, that is centered at Gd$^{3+}$ explains the exaggerated plateau feature that is observed using method (1).

\textit{Method 2: Height ratio}\\
Method (2) is the most straightforward and works well for fingerprinting when minimal plural scattering is evident in a spectrum. It is sensitive to the predominant electron energy-loss event, but does not account for the other similar electron energy-loss processes (the spread of the peak). When plural scattering is present, the M$_4$ edge height is increased more than the M$_5$ edge height. Because the multiplicity of the \textit{3d}10\textit{4f}n$\rightarrow $\textit{3d}9\textit{4f}n+1 transitions are not considered with the subtraction of a 3:2 step function, the M$_4$ edge is favored. The effects of plural scattering and \textit{3d} electron degeneracy are not accounted for with this method, resulting in an over-assessment of the M$_4$ edge height relative to the M$_5$ edge height.

While method (2) does not account for energy level multiplicity, plural scattering, or a range of energy-loss events near that of the edge maxima, it does offer the benefit of simplicity. After background subtraction, the relative edge maxima can be measured and compared to those of known RE-oxides. The method also does not require approximation of edge onset and end. However, as RE-oxide’s spectra remain relatively understudied, it is difficult to find standardized spectra for the various valencies of some of the RE-ions. 

Despite the deficiencies of the method, as is evident in Figure \ref{fig:fig3}, it should be noted that the edge height ratio trend as a function of \textit{4f} electron occupancy closely follows those of the edge height ratio with step function subtraction and edge second derivative methods. Predictably, the M$_4$/M$_5$ values for the edge height ratio without step function subtraction are always slightly greater than those for the edge height ratio with step function subtraction method because the unaccounted for 3:2 ratio of initial electron states.

\textit{Method 3: Height ratio with step function subtraction}\\
Method (3) is rarely used as the edge height ratio without step function subtraction allows for quicker RE-oxide fingerprinting. It has been included to show its similarity in trend with the other methods and for a comparison with the edge area method that also accounts for initial energy state multiplicity.

Both methods (2) and (3) show a plateau in M$_4$/M$_5$ feature ratio surrounding Gd$^{3+}$ However, these two methods also both indicate slight decrease in the M$_4$/M$_5$ feature ratio at Gd$^{3+}$ itself. It is likely that the depression results from the methods being entirely insensitive to the energy spread of the edges. Figure \ref{fig:fig4} shows a sharp increase in the FWHM of the M$_4$ edge and a decrease in the FWHM of the M$_5$ edge of Gd$^{3+}$. The edge height with step function subtraction can be applied for fingerprinting and when only the predominant (most intense) M$_{4,5}$  energy-loss transitions are desired for analysis. Method (3) shows a trend similar to the edge second derivative method, but, by neglecting the energy spread of the edges, it does not provide any information on the complexities of the energy-loss interactions happening during a \textit{3d}$\rightarrow $\textit{4f} electron event.

\textit{Method 4: Second derivative ratio}\\
Method (4) is performed independently of the confounding spectra contributions of plural scattering, \textit{3d} energy multiplicity, and the background energy continuum. This does not mean they are accounted for, just that they may be disregarded during analysis. It is the most robust method of calculating white line intensity ratios. It provides an easy method of fingerprinting white line spectra, but does not directly yield information on the energy spread of the electron. Because it is calculated front the rate of change of the slopes of the M$_{4,5}$ edges it includes some measure of spread, but not directly as the edge area ratio method does. The slope and geometry of the white lines influence the second derivative measurements. Beyond fingerprinting, this method is best applied when the M$_4$ and M$_5$ edges have similar slope and concavity. The M$_4$/M$_5$ trend as a function of \textit{4f} electron occupancy for method (4)  closely follows that of method (3). The greatest deviation between the two occurs for the later trivalent RE ions where the second derivative yields a lower M$_4$/M$_5$ feature ratio.

\end{document}